\documentclass[intlimits,twoside,a4paper]{article}

\usepackage{amsmath,amssymb}
\usepackage{graphicx}
\usepackage{wrapfig}

\usepackage[T2A]{fontenc}
\usepackage[cp1251]{inputenc}

\usepackage{cmpj2}

\issue{2013}{16}{3}{33701}
\doinumber{10.5488/CMP.16.33701}

\title[Optimization of quantum cascade laser operation ]%
{Optimization of quantum cascade laser operation by geometric
design of cascade active band in open and closed models}

\author[M.V.~Tkach \textsl{et al.}]{M.V.~Tkach\thanks{E-mail: ktf@chnu.edu.ua}\ , Ju.O.~Seti, I.V.~Boyko, O.M.~Voitsekhivska}
\address{Chernivtsi National University, 2 Kotsyubinsky St.,
58012 Chernivtsi, Ukraine}

\date{Received November 20, 2012}
\authorcopyright{M.V.~Tkach, Ju.O.~Seti, I.V.~Boyko, O.M.~Voitsekhivska, 2013}

\begin{document}

\maketitle

\begin{abstract}

Using the effective mass and rectangular potential approximations,
the theory of electron dynamic conductivity is developed for the
plane multilayer resonance tunnel structure placed into a
constant electric field within the model of open nanosystem, and
oscillator forces of quantum transitions within the model of
closed nanosystem. For the experimentally produced quantum cascade
laser with four-barrier active band of separate cascade, it is
proven that just the theory of dynamic conductivity
in the model of open cascade most adequately describes the
radiation of high frequency electromagnetic field while the
electrons transport through the resonance tunnel structure driven
by a constant electric field.
\pacs 73.40.Gk, 85.30.Mn, 81.07.St
\end{abstract}

\section{Introduction}

It is well known that the operation of quantum cascade laser (QCL)
\cite{1,2,3,4}, quantum cascade detector (QCD) \cite{5,6,7} and other appropriately
operating nanodevices is based on the transport properties of
open multilayer nanostructures. In spite of the long period of
investigations into electron transport through the resonance
tunnel structures (RTS), taking into account the interaction of
electronic current with constant electric and high frequency
electromagnetic fields, the current theory is still far from being well
correlated with experimental data.

The main problems in elaborating a consistent theory of electron transport
through the RTS are the mathematical difficulties arising when solving
the non-stationary Schrodinger equation with Hamiltonians
for even comparatively simple models with open boundaries which allow
the  infinite movement of quasi-particles. In order to avoid these
difficulties, the evaluations obtained for the closed analogues of
open RTS with rectangular potential wells and barriers were used
in early papers \cite{2,3,4} for the theory of electron transport
through the QCL active bands. The closed models did not make it possible
to study the currents due to the stationary electron states but
they satisfactorily described the electron spectrum and, thus, the
energies of electromagnetic radiation and wave functions which were
used for the calculation of dipole moments of quantum transitions.

Active bands of QCL, such as open two- and three-barrier RTS, were
theoretically studied in \cite{8,9,10}. In these papers, the
non-stationary one-dimensional Schrodinger equation describing the
electron transport through the RTS with $\delta$-like potential
barriers was solved taking into account the interaction with
constant electric and high frequency electromagnetic fields. The
simplified model of a constant effective mass of an electron along the
whole nanosystem and $\delta$-barrier approximation of the
potential made it possible to calculate and investigate the
electronic currents and, consequently, to calculate the dynamic conductivity in
ballistic regime when the biggest lifetimes in operating
quasi-stationary states were much smaller than the relaxation times
for the electron energy due to the dissipative processes (phonons,
impurities and so on).

The $\delta$-barrier model for an open RTS \cite{11,12} essentially
overestimates the resonance widths of operating quasi-stationary
states compared to the more adequate model of rectangular
potentials. It explained some properties of electron transport
but could not be used as a reliable base to be compared with
the experimental data and, thus, to optimize the geometric
design of QCL active band.

According to the abovementioned and using the effective mass and
rectangular potential approximations, in this paper we develop
the theory of quasi-stationary spectrum and dynamic
conductivity of the electrons interacting with high frequency
electromagnetic field within the model of open multilayer RTS and
stationary spectrum together with oscillator forces of quantum
transitions within the model of closed RTS in a constant electric
field. We use the obtained theoretical results in order to
calculate the energy of electromagnetic radiation for the
experimentally produced QCL \cite{3} with a four-barrier active band of
a separate cascade. The comparison of numerical and experimental
data illustrates the capabilities of different models in
optimizing  the active band geometric design.

\section{Theory of dynamic conductivity of a resonance tunnel cascade with four-barrier active band and oscillator forces of quantum transitions in closed model}

\begin{figure}[!b]
\centerline{
\includegraphics[width=0.6\textwidth]{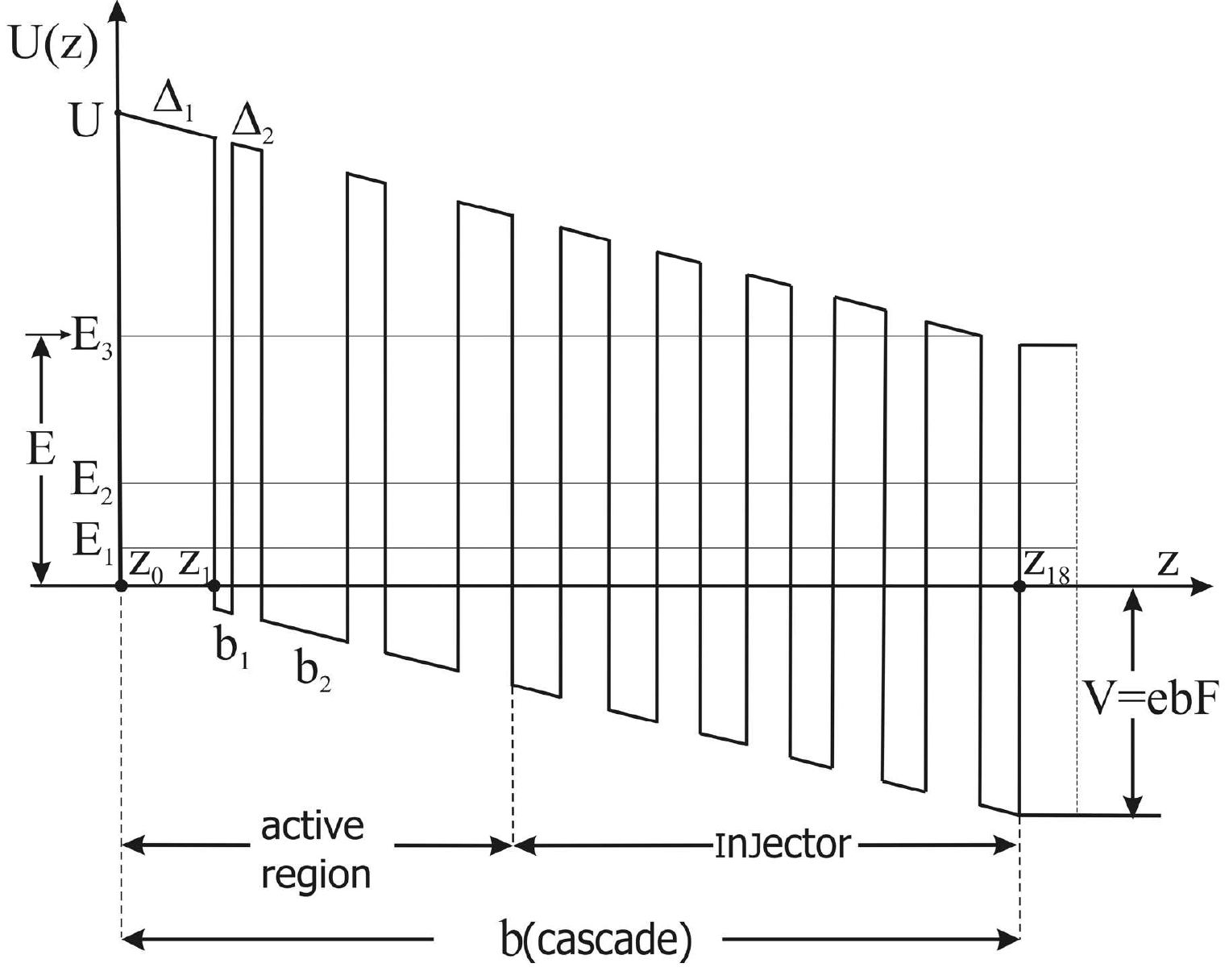}
}
\caption{The energy scheme of separate cascade with four-barrier
active region and injector. The widths of the barriers
($\Delta_{p}$): 5.0, 1.5, 2.2, 3.0, 2.3, 2.2, 2.0, 2.3, 2.8 and
widths of the wells ($b_{p}$): 0.9, 4.7, 4.0, 2.3, 2.2, 2.0, 2.0,
1.9, 1.9 are presented from the left to the right in nm units.}
\label{fig1}
\end{figure}

The separate cascade of QCL such as RTS, containing a four-barrier
active band and injector consisting of a certain number of plane
nanolayers (wells and barriers) having fixed sizes, figure~\ref{fig1}, is
studied within two models: open (o) and closed (c). The constant
electric field with intensity F is applied perpendicularly to the
RTS planes. For the open model, we assume that the monoenergetic
current of non-interacting electrons having energy E and
concentration n0 impinges at RTS from the left hand side, perpendicularly
to its planes. Under these conditions and taking into account the
small difference between the lattice constants of wells and
barriers, the problem settles to the study of one-dimensional
electron transport using the models of effective mass and
rectangular potentials.

Taking the coordinate system as it is shown in figure~\ref{fig1}, the
effective mass and potential energy of an electron in open (o) or
closed (c) RTS (without the field) is conveniently written as
\begin{eqnarray}
\label{eq1}
m_{\left\{\begin{smallmatrix}\textrm{o}\\\textrm{c}\end{smallmatrix}\right\}} (z)&=&\left\{\begin{array}{l} {m_{0} } \\
{m_{1} }
\end{array}\right\}[\theta (-z)+\theta (z-b)]+m_{0} \sum
_{p=1}^{N_\mathrm{W} }\left[\theta (z-z_{2p-1} )-\theta (z-z_{2p}
)\right]\nonumber\\
& + & m_{1} \sum _{p=0}^{N_\mathrm{B} -1}\left[\theta (z-z_{2p} )-\theta
(z-z_{2p+1} )\right],
\end{eqnarray}
\begin{equation}
\label{eq2} U{}_{\left\{\begin{smallmatrix}\textrm{o}\\\textrm{c}\end{smallmatrix}\right\}} (z)=\left\{\begin{array}{l} {0} \\ {U}
\end{array}\right\}[\theta (-z)+\theta (z-b)]+U\sum _{p=0}^{N_\mathrm{B}
-1}\left[\theta (z-z_{2p} )-\theta (z-z_{2p+1} )\right],
\end{equation}
where $N_\mathrm{W}$,  $N_\mathrm{B}$ are the numbers of wells and barriers in
the RTS which correspond to the active band or to the whole cascade,
depending on the model.

In order to observe the electron transport through the RTS, the
latter should be obligatory an open one. Thus, a dynamic
conductivity arises when quasi-stationary states are present in
a nanosystem. Developing the theory of RTS active conductivity
within the open model, we study the properties of electron stationary
spectrum and oscillator forces of quantum transitions within the closed
model in order to be compared. The widths of outer barriers of active band
(or cascade) in the closed model limit to the physical infinity and the
constant electric field is applied inside the nanosystem only. The
reason to study the closed model is that the similar model is, evidently,
the theoretical base of the choice of experimental geometric design
of QCL cascade with active band and injector \cite{3}. We are going to compare
the experimental data with our results obtained within the model of open RTS.

In order to calculate the dynamic conductivity within the
open model and oscillator forces within the closed one, we
first solve the stationary Schrodinger equations
\begin{equation}
\label{eq3} H_{\left\{\begin{smallmatrix}\textrm{o}\\\textrm{c}\end{smallmatrix}\right\}} (z)\Psi _{\left\{\begin{smallmatrix}\textrm{o}\\\textrm{c}\end{smallmatrix}\right\}} (z)=E\Psi _{\left\{\begin{smallmatrix}\textrm{o}\\\textrm{c}\end{smallmatrix}\right\}} (z)
\end{equation}
with the Hamiltonian of an electron in RTS driven by a constant
electric field
\begin{equation}
\label{eq4} H_{_{\left\{\begin{smallmatrix}\textrm{o}\\\textrm{c}\end{smallmatrix}\right\}} } =-\frac{\hbar ^{2} }{2} \frac{\partial
}{\partial z} m_{_{\left\{\begin{smallmatrix}\textrm{o}\\\textrm{c}\end{smallmatrix}\right\}} }^{-1} (z)\frac{\partial }{\partial z}- e F{\left( {z[\theta (z) - \theta (z - b)]- {\left\{ \begin{array}{l} {b } \\
{0} \end{array}\right\}}\theta (z - b)} \right)}.
\end{equation}
The solutions of equations (\ref{eq3}) are written as follows:
\begin{eqnarray}
\Psi _{\left\{\begin{smallmatrix}\textrm{o}\\\textrm{c}\end{smallmatrix}\right\}} (z)&=&\Psi _{\left\{\begin{smallmatrix}\textrm{o}\\\textrm{c}\end{smallmatrix}\right\}}^{(0)} (z)\theta (-z)+\sum _{p=1}^{N_\mathrm{W}
+N_\mathrm{B} }\Psi _{\left\{\begin{smallmatrix}\textrm{o}\\\textrm{c}\end{smallmatrix}\right\}}^{(p)} (z)\left[\theta (z-z_{p-1} )-\theta
(z-z_{p} )\right]\nonumber\\
\label{eq5}
& + &  \Psi _{\left\{\begin{smallmatrix}\textrm{o}\\\textrm{c}\end{smallmatrix}\right\}}^{(N_\mathrm{W} +N_\mathrm{B} +1)} (z)\theta (z-b),
\end{eqnarray}
where the wave functions
\begin{eqnarray}
\label{eq6} \Psi _{\left\{\begin{smallmatrix}\textrm{o}\\\textrm{c}\end{smallmatrix}\right\}}^{(0)} (z)&=&A_{\left\{\begin{smallmatrix}\textrm{o}\\\textrm{c}\end{smallmatrix}\right\}}^{(0)} \mathrm{e}^{\left\{\begin{smallmatrix} {\mathrm{i}k} \\ {\chi } \end{smallmatrix}\right\}z} +B_{\left\{\begin{smallmatrix}\textrm{o}\\\textrm{c}\end{smallmatrix}\right\}}^{(0)} \mathrm{e}^{-\left\{\begin{smallmatrix} {\mathrm{i}k} \\ {\chi } \end{smallmatrix}\right\}z}, \\[2ex]
%
\label{eq7} \Psi _{\left\{\begin{smallmatrix}\textrm{o}\\\textrm{c}\end{smallmatrix}\right\}}^{(p)} (z)&=&A_{\left\{\begin{smallmatrix}\textrm{o}\\\textrm{c}\end{smallmatrix}\right\}}^{(p)}{\rm A}{\rm
\mathrm{i}}(\xi _{p} (z)) +B_{\left\{\begin{smallmatrix}\textrm{o}\\\textrm{c}\end{smallmatrix}\right\}}^{(p)} {\rm B}{\rm \mathrm{i}}(\xi _{p}
(z)), \qquad [p = 1\div (N_\mathrm{W} + N_\mathrm{B})], \\
%
\label{eq8} \Psi _{\left\{\begin{smallmatrix}\textrm{o}\\\textrm{c}\end{smallmatrix}\right\}}^{(N_\mathrm{W} +N_\mathrm{B} +1)}
(z)&=&A_{\left\{\begin{smallmatrix}\textrm{o}\\\textrm{c}\end{smallmatrix}\right\}}^{(N_\mathrm{W} +N_\mathrm{B} +1)}
\mathrm{e}^{\left\{\begin{smallmatrix} {\mathrm{i}K} \\ {\chi} \end{smallmatrix}\right\}z} +B_{\left\{\begin{smallmatrix}\textrm{o}\\\textrm{c}\end{smallmatrix}\right\}}^{(N_\mathrm{W} +N_\mathrm{B} +1)}
\mathrm{e}^{-\left\{\begin{smallmatrix} {\mathrm{i}k} \\ {\chi} \end{smallmatrix}\right\}z}
\end{eqnarray}
are the superpositions of the exact linearly independent
solutions of equations (\ref{eq3}) in the respective ranges of $z$ variable.
Here, we introduced the notations
\begin{align}
k = \hbar ^{ - 1}\sqrt {2m_{0}
E}, \qquad \chi &= \hbar ^{ - 1}\sqrt {2m_{1} (U - E)},\qquad K = \hbar ^{ - 1}\sqrt {2m_{0} (E + V)}, \qquad V = eFb, \nonumber\\[2ex]
\xi _{p} (z) &= {\left\{ {\begin{array}{l}
 \displaystyle { + \left( {{\frac{{2m{}_{1}\,V\,b^{2}}}{{\hbar ^{2}}}}}
\right)^{{\frac{{1}}{{3}}}}\left( {{\frac{{U - E}}{{V}}} -
{\frac{{z}}{{b}}}} \right), \qquad p = 1,\,3,\,5,\,\ldots \ , } \\
 \displaystyle{ - \left( {{\frac{{2m{}_{0}\,V\,b^{2}}}{{\hbar ^{2}}}}}
\right)^{{\frac{{1}}{{3}}}}\left( {{\frac{{E}}{{V}}} - {\frac{{z}}{{b}}}}
\right), \qquad p = 2,\,4,\,6,\,\ldots \ .} \\
 \end{array}}\ \right.}
 \end{align}
 $\textrm{Ai}(\xi ),\,\textrm{Bi}(\xi )\,$ are the Airy functions.

The conditions of a wave function and its density of current
continuity should be fulfilled at all nanosystem interfaces in the both
models
\begin{equation}
\label{eq10} \Psi _{\left\{\begin{smallmatrix}\textrm{o}\\\textrm{c}\end{smallmatrix}\right\}}^{(p)} (z_{p} )=\Psi _{\left\{\begin{smallmatrix}\textrm{o}\\\textrm{c}\end{smallmatrix}\right\}}^{(p+1)} (z_{p} ),
\qquad \left. \frac{\rd\Psi _{\left\{\begin{smallmatrix}\textrm{o}\\\textrm{c}\end{smallmatrix}\right\}}^{(p)} (z)}{m_{\left\{\begin{smallmatrix}\textrm{o}\\\textrm{c}\end{smallmatrix}\right\}} (z)\rd z} \right|_{z=z_{p}-\varepsilon} =\left. \frac{\rd \Psi _{\left\{\begin{smallmatrix}\textrm{o}\\\textrm{c}\end{smallmatrix}\right\}}^{(p+1)} (z)}{m_{\left\{\begin{smallmatrix}\textrm{o}\\\textrm{c}\end{smallmatrix}\right\}} (z)\rd z} \right|_{z=z_{p}+\varepsilon} ,
\end{equation}
$p=0 \div  (N_\mathrm{W} +N_\mathrm{B})$, $\varepsilon \to +0$.

The wave functions tend to zero at $z\rightarrow±\infty$
in the closed model, since $B_{(\mathrm{c})}^{(0)} =A_{(\mathrm{c})}^{(N_\mathrm{W} +N_\mathrm{B}
+1)} =0.$   Thus, the system of equations (\ref{eq10}) brings us to the
dispersion equation, consistently determining the energy spectrum
($E_{n}$) and all coefficients $A_{(\mathrm{c})}^{(p)} ,\,\,B_{(\mathrm{c})}^{(p)}$
through one of them. The latter is obtained from the normality
condition
\begin{equation} \label{eq11}
{\int\limits_{ - \infty} ^{\infty}  {\Psi _{(\mathrm{c})n}^{ *}  (z)\Psi
_{(\mathrm{c}){n}'}} }(z)\rd z = \delta _{n{n}'} \,.
\end{equation}

Now, the electron wave functions $\Psi _{(\mathrm{c}){n}}(z)$  and energies
($E_{n}$) of all stationary states are defined in the closed model.
Using them, the oscillator forces of quantum transitions
between the states  $n$  and  ${n}'$ can be calculated within the
formula
\begin{equation} \label{eq12}
f_{n{n}'} = {\frac{{2(E_{n} - E_{{n}'} )}\overline{m}_{(\mathrm{c})}}{{\hbar ^{2}}}}{\left|\,
{{\int\limits_{ - \infty} ^{\infty}  { \Psi _{(\mathrm{c})n}^{ *}
(z)\,z\,\Psi _{(\mathrm{c}){n}'} (z)\rd z}} } \right|}^{2}.
\end{equation}

For the open model, there should be no backward wave from the right of
a nanosystem, since \linebreak $B_{(\mathrm{o})}^{(N_\mathrm{W} + N_\mathrm{B} + 1)} =
0$. All coefficients $A_{(\mathrm{o})}^{(p)}$, $B_{(\mathrm{o})}^{(p)}$   of the
wave function  $\Psi _{(\mathrm{o})} (z)$  are found from the condition (\ref{eq10})
through one of them, in its turn, defined by the incident density
of the current impinging at RTS from the left hand side. In this case, the
electron spectrum is the quasi-stationary one with the resonance
energies ($E_{n}$) and resonance widths ($\Gamma_{n} =
\hbar \tau _{n}^{ - 1}$) where $\tau_{n}$  is the lifetime in the
$n$-th quasi-stationary state. The resonance energies are fixed by
the maxima of a probability distribution function of an electron inside
RTS (in energy scale $E$)
\begin{equation} \label{eq13}
W(E) = {\frac{{1}}{{b}}}{\int\limits_{0}^{b} {\,\,{\left| {\Psi
_{(\mathrm{o})} (E,\,z)} \right|}^{2}\rd z}}.
\end{equation}
The resonance widths $(\Gamma_{n})$ are fixed by the widths of
this function at the halves of its maxima placed at the
respective resonance energies $E_{n}$.

The quantum transitions between the quasi-stationary states occur when
the electrons transport through the open RTS placed into the
electric field. Consequently, the electromagnetic field with the respective frequency arises. Its intensity is proportional to the
magnitude of the dynamic conductivity. In the quantum transitions
accompanied by the absorption of electromagnetic energy, the positive
dynamic conductivity is formed and, during the radiation of
electromagnetic energy, the negative dynamic conductivity of RTS
is formed.

In order to calculate the negative conductivity of open RTS
operating in a laser regime, one has to obtain the wave functions of
electrons interacting with the electromagnetic field. It is found
from the time-dependent Schr\"{o}dinger equation:
\begin{equation} \label{eq14}
\mathrm{i}\hbar {\frac{{\partial \Psi (z,t)}}{{\partial t}}} =
\left[ {H_{(\mathrm{o})} (z) + H(z,t)} \right]\,\,\Psi (z,t),
\end{equation}
where $H_{(\mathrm{o})} (z)$   is the Hamiltonian (\ref{eq4}) for the electrons in
RTS without the electromagnetic field and
\begin{equation} \label{eq15}
H(z,t) = - e\mathcal{E}{\left\{ {z\,\left[ {\theta \left( {z}
\right) - \theta \left( {z - b} \right)} \right] + b\,\theta
\left( {z - b} \right)} \right\}}{\rm }\left( {
\mathrm{e}^{\mathrm{i}\omega t} + \mathrm{e}^{ - \mathrm{i}\omega
t}} \right)
\end{equation}
is the Hamiltonian of electrons interacting with time-dependent
electromagnetic field characterized by the frequency $\omega$
and its electric field intensity $\mathcal{E}$.

Assuming the amplitude of a high frequency electromagnetic field to be small, we find the solution of equation (\ref{eq14}) in a one-mode
approximation using the perturbation theory
\begin{equation} \label{eq16}
\Psi \left( {z,t} \right) = {\sum\limits_{s\, = - 1}^{ + 1} {\Psi
_{s} \left( {z} \right)\,\mathrm{e}^{ - \mathrm{i}\left( {\omega
_{0} + s\omega} \right) t}} }, \qquad (\omega _{0} =
{{E}/{\hbar}}),
\end{equation}
where $\Psi _{s= 0} (z) \equiv \Psi_{(\mathrm{o})}(z)$.

Preserving the first order magnitudes in equation (\ref{eq14}), we obtain
inhomogeneous equations for the corrections $\Psi _{\pm 1}(z)$ to the wave functions
\begin{equation} \label{eq17}
\displaystyle {\left[ {H_{(\mathrm{o})} \left( {z} \right) - \hbar \left( {\omega _{0}
\pm \omega} \right)} \right]}{\rm} \Psi _{\pm 1} \left( {z}
\right) - e\mathcal{E}{\left\{ {z\,\left[ {\theta \left( {z}
\right) - \theta \left( {z - b} \right)} \right] + b\,\theta
\left( {z - b} \right)} \right\}}{\rm} \Psi _{0} \left( {z}
\right) = 0.
\end{equation}
Their solutions are the superpositions of functions
\begin{equation} \label{eq18}
\Psi _{\pm 1} \left( {z} \right) = \Psi _{\pm}  \left( {z} \right)
+ \Phi _{\pm}  \left( {z} \right).
\end{equation}

The functions $\Psi _{\pm}  \left( {z} \right)$, being the solutions
of homogeneous equations, are written as
\begin{eqnarray} \label{eq19}
 \Psi _{\pm}  (z) &=& \Psi _{\pm} ^{(0)} (z)\theta ( -
z) + {\sum\limits_{p = 1}^{N_\mathrm{W} + N_\mathrm{B}}  {\Psi _{\pm} ^{(p)}
(z){\left[ {\theta (z - z_{p - 1} ) - \theta (z - z_{p} )}
\right]}}}  + \Psi _{\pm} ^{(N_\mathrm{W} + N_\mathrm{B} + 1)} (z)\theta (z - b)
\nonumber \\
& = & B_{\pm} ^{(0)} \mathrm{e}^{ - \mathrm{i}k_{\pm}  z}\theta ( - z) + A_{\pm} ^{(N_\mathrm{W} + N_\mathrm{B}
+ 1)} \mathrm{e}^{\mathrm{i}K_{\pm}  z}\theta (z - b)  \nonumber \\
&+ & {\sum\limits_{p = 1}^{N_\mathrm{W} + N_\mathrm{B}}  {\left[ {A_{\pm} ^{(p)} {\rm A}{\rm
\mathrm{i}}(\xi _{\pm} ^{(p)} ) + B_{\pm} ^{(p)} {\rm B}{\rm
\mathrm{i}}(\xi _{\pm} ^{(p)} )} \right]\,{\left[ {\theta (z -
z_{p - 1} ) - \theta (z - z_{p} )}
\right]}{\rm} {\rm} }} {\rm} ,
\end{eqnarray}
where

\[
k_{\pm}  = \hbar ^{ - 1}\sqrt {2m_{0} (E\pm \Omega )}, \qquad K_{\pm}  = \hbar ^{ - 1}\sqrt
{2m_{0} [(E\pm \Omega ) + V]}, \qquad \Omega = \hbar
\omega,
\]
\begin{eqnarray}
\label{eq20}
\xi _{\pm} ^{(p)} (z) =  {\left\{ {\begin{array}{l}
 {\displaystyle + \left( {{\frac{{2m{}_{1}\,V\,b^{2}}}{{\hbar ^{2}}}}}
\right)^{{\frac{{1}}{{3}}}}\left[ {{\frac{{U - (E\pm \Omega )}}{{V}}} -
{\frac{{z}}{{b}}}} \right], \qquad p = 1,\,3,\,5,\,\ldots \ ,} \\
{\displaystyle - \left( {{\frac{{2m{}_{0}\,V\,b^{2}}}{{\hbar ^{2}}}}}
\right)^{{\frac{{1}}{{3}}}}\left( {{\frac{{E\pm \Omega} }{{V}}} -
{\frac{{z}}{{b}}}} \right), \qquad p =
2,\,4,\,6,\,\ldots \ .} \\
 \end{array}}\ \right.}
\end{eqnarray}

The partial solutions of inhomogeneous equations (\ref{eq17}) have the exact analytical form
\begin{eqnarray} \label{eq21}
\Phi _{\pm}  (z) &=& \pi {\frac{{\mathcal{E}}}{{F}}}{\sum\limits_{p = 1}^{N_\mathrm{W} + N_\mathrm{B}
} {{\left\{ {{\rm B}{\rm i}(\xi _{\pm} ^{(p)} ){\int\limits_{1}^{\xi ^{(p)}}
{{\left[ {\eta - \kappa^{{\frac{{2}}{{3}}}}(z)\,{\frac{{U(z) - E}}{{V}}}}
\right]}{\rm A}{\rm i}\left( {\eta \mp
\kappa^{{\frac{{2}}{{3}}}}(z)\,\,{\frac{{\Omega} }{{V}}}} \right)\Psi
_{(\mathrm{o})}^{(p)} \left( {\eta}  \right)\,\textrm{d}\eta  \,  \,}} } \right.}}} \nonumber \\
&-&{\left. { {\rm A}{\rm i}(\xi _{\pm} ^{(p)} ){\int\limits_{1}^{\xi ^{(p)}}
{{\left[ {\eta - \kappa^{{\frac{{2}}{{3}}}}(z)\,\,{\frac{{U(z) - E}}{{V}}}}
\right]}{\rm B}{\rm i}\left( {\eta \mp
\kappa^{{\frac{{2}}{{3}}}}(z)\,\,\,{\frac{{\Omega} }{{V}}}} \right)\Psi
_{(\mathrm{o})}^{(p)} \left( {\eta}  \right)\,\textrm{d}\eta} } } \right\}}   \\
\, \nonumber \\
&\times & {\left[ {\theta (z
- z_{p - 1} ) - \theta (z - z_{p} )} \right]} \ \mp \ {\frac{{e\mathcal{E}b}}{{\Omega} }}\Psi _{(\mathrm{o})}^{(N_\mathrm{W} + N_\mathrm{B} + 1)} \left( {b}
\right){\rm} \theta (z - b){\rm},
\end{eqnarray}
where
\begin{equation} \label{eq22}
\kappa(z) =\hbar ^{-1} \sqrt {2m_{(\mathrm{o})}(z)b^{2}V}.
\end{equation}

The conditions of the wave function $\Psi (z,t)$ and its density of current
continuity at all RTS interfaces bring us to the fitting conditions similar to the equations (\ref{eq10}) for the functions $\Psi _{\pm 1} \left( {z} \right)$.
Also, these equations define the unknown coefficients $B_{\pm} ^{(p)}$, $A_{\pm} ^{(p)}$ [$p = 0\div ({\rm N}_{{\rm W}} + {\rm
N}_{{\rm B}} + {\rm 1}{\rm )}{\rm}$], and, consequently, the complete wave function $\Psi (z,t)$.

Further, considering the energy of electron-electromagnetic field interaction
to be the sum of energies of electron waves, coming out of the both sides of RTS,
we calculate, in quasi-classic approximation, the real part of dynamic conductivity
through the densities of currents of electron waves coming out of the both sides of a nanosystem
\begin{equation} \label{eq22.2}
\sigma (\Omega ,E) = {\frac{{\Omega} }{{2be\mathcal{E}^{2}}}}{\Big\{ {{\rm} {\left[
{j\left( {E + \Omega ,z = b} \right) - j\left( {E - \Omega ,z = b} \right)}
\right]}}  { - {\left[ {j\left( {E + \Omega ,z = 0} \right) -
j\left( {E - \Omega ,z = 0} \right)} \right]}{\rm} } \Big\} } {\rm .}
\end{equation}
According to the quantum mechanics, the densities of currents are determined by the wave function
\begin{equation} \label{eq23}
j(E,z) = {\frac{{\mathrm{i}e\hbar n_{0}} }{{2m_{(\mathrm{o})} (z)}}}{\left[
{\Psi (E,z){\frac{{\partial} }{{\partial z}}}\Psi ^{\ast} (E,z) -
\Psi ^{\ast }(E,z){\frac{{\partial} }{{\partial z}}}\Psi (E,z)}
\right]}.
\end{equation}

The real part of dynamic conductivity can be expressed as a sum of two terms
\begin{eqnarray} \label{eq24}
&& \sigma ^{ -} (\Omega ,E) = {\frac{{\hbar \,\Omega {\rm} n_{0}} }{{2b\,m_{0}
\mathcal{E}^{2}}}}\left( {k_{ +}  {\left| {B_{ +} ^{(0)}}  \right|}^{2} - k_{ -}
{\left| {B_{ -} ^{(0)}}  \right|}^{2}} \right),\, \nonumber\\
&& \sigma ^{ +} (\Omega ,E) = {\frac{{\hbar \,\Omega {\rm} n_{0}} }{{2b\,m_{0}
\mathcal{E}^{2}}}}\left( {K_{ +}  {\left| {A_{ +} ^{(N_\mathrm{W} + N_\mathrm{B} + 1)}}
\right|}^{2} - K_{ -}  {\left| {A_{ -} ^{(N_\mathrm{W} + N_\mathrm{B} + 1)}}
\right|}^{2}} \right).
\end{eqnarray}
The physical sense of these partial terms [$\sigma ^{\pm} (\Omega ,E)$] is evident.
They are caused by the electronic currents interacting with high frequency electromagnetic
field in RTS and flowing out of it in forward ($\sigma ^{ +}$) and backward ($\sigma ^{ -}$) direction
with respect to the incident one.

\section{Discussion of the results}

Using the developed theory we display such a model of plane
nano-RTS, which describes the quantum transitions and transport
properties of electrons best of all. It makes possible to
optimize the operation of QCL by the geometric design of separate
cascade active band.  The numeric calculations were performed for
four nanosystems: two closed models(four-barrier active
band and complete cascade)and two open models(four-barrier
active band and complete cascade).

In order to compare with the experimental data \cite{3} we used the following
physical parameters: $U=516$~meV, $F = 68$~kV/cm,
$n_{0} = 2 \cdot 10^{17}$~cm$^{ - 3}$ and geometrical ones, shown in
figure 1. We should note that as far as almost equal sizes of all
layers in the experimentally investigated cascade in the cited
paper contain a small number of unitary cells (2--4) of its
composition elements, the approximation of effective masses
in different layers of a nanosystem would be a rough one. At the
same time, the whole active band, the whole injector or the whole
cascade contains dozens of unitary cells in composition elements.
Thus, one can expect that the effective mass of an electron
($m=0.08 \, m_{e}$)averaged over all three composition elements
(GaAs, AlAs, InAs) is more adequate in the present physical
situation.

\begin{figure}[!t]
\centerline{
\includegraphics[width=0.8\textwidth]{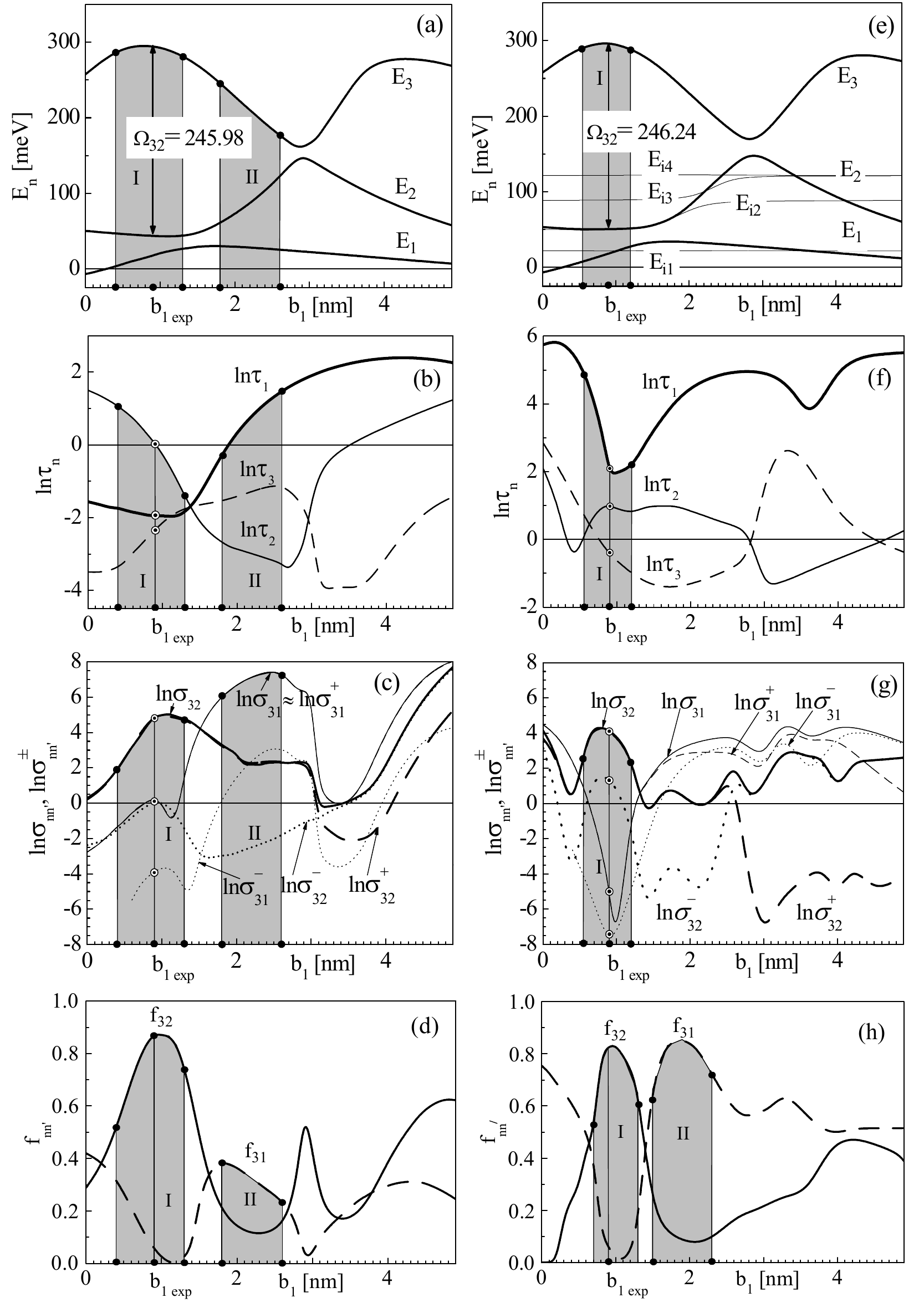}
}
\caption{Electron energy spectrum (a, e), lifetimes (b, f),
oscillator forces (d, h) and active conductivities (c, g) as
functions of an input well width ($b_{1}$) of QCL active region in
different models.} \label{fig2}
\end{figure}

In order to study the effect of a geometric design of a separate
cascade on the operation of QCL we calculated the energy spectrum
($E_{n}$) and oscillator forces of quantum transitions
($f_{n n'}$) within the closed model, while resonance energies ($E_{n}$),
lifetimes ($\tau_{n}$), active conductivity ($\sigma_{nn'}$) and
its partial terms ($\sigma _{n{n}'}^{\pm}$)~--- within the open model. The
results are presented in figure~\ref{fig2}, depending on the position
($b_{1}$) of inner two-barrier element between two outer barriers
of an active band at the fixed sizes of all other elements of
a cascade, the same as in paper \cite{3}.

The calculations prove that the dependences of electron spectra on
$b_{1}$ in closed and open systems differ not more than by
0.1$\%$. Besides, from figures~\ref{fig2}~(a), (e) it is clear that the first three
resonance energies, as functions of $b_{1}$, calculated within the
model of an open four-barrier active band [figure~\ref{fig2}~(a)] coincide with the
respective resonance energies ($E_{1}$, $E_{2}$, $E_{3}$) of those
three states, calculated within the model of an open complete cascade
[figure~\ref{fig2}~(e)] where the electron with maximal probability is located in
the space of an active band. In figure~\ref{fig2}~(e) one can also see the resonance
energies $E_{i1}\div E_{i4}$ (thin curves) of quasi-stationary
states, where the electron with a bigger probability is located in
the injector part of a cascade.

Using the closed model, we calculated the oscillator forces of
quantum transitions ($f_{32}$ and $f_{31}$) as functions of
$b_{1}$. The results are presented for the four-barrier active
band [figure~\ref{fig2}~(d)] and for the complete cascade [figure~\ref{fig2}~(h)]. The condition
of optimal QCL operation is fulfilled when the oscillator force of
quantum transition $f_{32}$ approaches its maximal magnitude. This
transition occurs between the states ensuring the needed energy of
electromagnetic field radiation (in our case
$\Omega_{32}=E_{3}-E_{2}$). Herein, the oscillator force of
quantum transition $f_{31}$ should be an order smaller. Figure~\ref{fig2}~(d)
proves that for the four-barrier active band model, one can see two
regions (I and II) of $b_{1}$ varying [toned at the figure~\ref{fig2}~(d)] where
the abovementioned condition fulfills. Herein, the I region is
better than the II region because $f_{32}^{I} > f_{32}^{II}$. It is
clear that the experimental point $b_{1\mathrm{exp}}$ nearly corresponds to
the maximal magnitude of $f_{32}$ as a function of $b_{1}$. In
a closed model of a complete cascade [figure~\ref{fig2}~(h)] there is one region of
$b_{1}$ varying (I) where $f_{32} > f_{31}$ and one region (II)
where $f_{31} > f_{32}$. Herein, the experimental point $b_{1\mathrm{exp}}$
also nearly coincides with $b_{1}$ corresponding to the maximal
magnitude of $f_{32}$. The width and location of an optimal region
for $b_{1}$ in this model are almost the same as in the model of
a four-barrier active band. Also, we should note that in the model of
a complete cascade it seems that the QCL  can operate due to the
quantum transition $ 3\to 1 $ in the II region where $f_{31} >
f_{32}$. However, as it is proven further on within the more
adequate open model, this is impossible.

The resonance energies ($E_{n}$, $E_{in}$), lifetimes
($\tau_{n}$), active conductivities ($\sigma_{32}$, $\sigma_{31}$)
and their partial terms ($\sigma _{32}^{\pm}  ,\,\sigma
_{31}^{\pm}$) are calculated as functions of $b_{1}$ for the two
open models: active band [figures~\ref{fig2}~(a), (b), (c)] and complete cascade
[figures~\ref{fig2}~(e), (f), (g)]. Before analysing the figures we should
note that in the model of an active band the dynamic conductivities and
their terms were calculated through the averaging over those
energy regions where the levels of an injector band ($E_{i4}-E_{i1}$) are uniformly located. The calculations of these
conductivities for the model of a separate complete cascade were
performed taking into account that the electrons leave the
previous cascade with the energy $E_{1}$, shifted at the magnitude
$E_{3} - E_{1} - eFb$ , with respect to the resonance energy
$E_{3}$ of the cascade studied.

Contrary to the closed model, the open one allows for
a detailed and adequate analysis of the conditions optimizing the QCL
operation due to the geometric design of an active band. Within the
open models, one can evaluate the magnitude of dynamic conductivity
in the needed quantum transition (for example, $ 3\to 2 $).
Besides, this conductivity would be much bigger than the
conductivity of the close over the energy transition (for example,
$ 3\to 1 $) under the condition that the partial term of conductivity
($\sigma _{32}^{+}$) in forward direction would be much bigger
than the partial term ($\sigma _{31}^{-}$) in backward current.
The calculated lifetimes ($\tau_{n}$) of the operating electron
quasi-stationary states make it possible to guide the natural
physical condition: these lifetimes should not exceed the
relaxation times of dissipative processes due to the scattering of
electrons at the impurities, phonons, imperfections of media
interfaces and other factors, which, according to the evaluations
\cite{4}, are not bigger than twenty picoseconds.

The results of numeric calculations of conductivities and lifetimes of electrons in the quasi-stationary states ($n = 1, \ 2, \ 3$)
obtained within the model of active band and complete cascade as
functions of $b_{1}$ are presented in figures~\ref{fig2}~(c), (b) and figures~\ref{fig2}~(g), (f),
respectively. Analysis of $\tau_{n}$, $\sigma _{n{n}'}$, $\sigma
_{n{n}'}^{\pm}$ dependences on $b_{1}$ [figures~\ref{fig2}~(b), (c)] proves that
there are two regions of $b_{1}$ for the model of a four-barrier
active band, where: I~--- the optimal is the quantum transition $
3\to 2 $ (experimentally observed), II~--- the optimal is the
quantum transition $ 3\to 1 $. The latter transition is possible
[figure~\ref{fig2}~(c)] the same as in the model of a closed cascade. In this
narrow region (II) for $b_{1}$, the lifetime in the first
quasi-stationary state is rather small ($\tau _{1} \leqslant 1\div 3\,\textrm{ps}
< 10\,\textrm{ps}$).

The model of an open complete cascade [figures~\ref{fig2}~(f), (g)] is the best one for
the description of the properties of an electron current through the
RTS of QCL with the electromagnetic radiation accompanying quantum
transitions. Figure~\ref{fig2}~(g) proves that in this model, the same as in the
model of an open active band, there are also two regions for $b_{1}$
varying (I and II) where the conditions of optimal QCL operation
fulfill well (the transitions $3\to 2 $  and $3\to 1 $,
respectively). The location and sizes of these regions are nearly
the same as in the model of an open active band and in the model of
a closed complete cascade.

However, the analysis of lifetimes $\tau_{1}$,  $\tau_{2}$,
$\tau_{3}$ [figure~\ref{fig2}~(f)] shows that in fact, the region II is not
optimal because at such a geometric design the lifetime  $\tau_{1}
\geqslant 10$~ps approaches the time of dissipative processes,
breaking the coherent regime of QCL. Thus, the model of an open
cascade proves that in the experimental QCL \cite{3} it is only one
narrow region (I) ($0.55~\textrm{nm} \leqslant b_{1}\leqslant 1.2~\textrm{nm}$) of the
position of the inner two-barrier structure between the outer
barriers of an active band, where the laser operates in an optimal regime.
Only this configuration ensures that the active conductivity
$\sigma_{32}$ in direct current is much bigger than the other
conductivities. Herein, not only the lifetimes of both operating
quasi-stationary states are small ($\tau_{3}, \tau_{2}\leqslant 2~\textrm{ps}$)
but the lifetime of the first quasi-stationary state
($\tau_{1}\leqslant 10~\textrm{ps}$) through which the electrons flow into
the next cascade due to the interaction with phonons \cite{3,4}, is
minimal.

The geometric and physical parameters for the numeric calculations
within four theoretical models were taken the same as in paper \cite{3}
in order to compare the theoretical and experimental data.
These parameters are presented in figure~\ref{fig1} and figure~\ref{fig2}. The numeric
calculations show that in all four models the energies ($E_{1}$,
$E_{2}$, $E_{3}$) of operating quasi-stationary states differ between
each other not more than by 0.1\%. Thus, in all models the
theoretical magnitude of the energy of laser radiation $\Omega
_{32} = E_{3} - E_{2} = 246$~meV differs from the experimental
one $\Omega _{32}^{\mathrm{exp}}  = 238.8$~meV by 3\%
and the difference of the energies $E_{3} - E_{2}=34$~meV nearly
coincides with the phonon energy in \cite{3}.

Finally, we should note that the experimental geometric design of
QCL cascade \cite{3} with the input well width $b_{1} = 0.9$~nm of a
four-barrier active band correlates well with the magnitude $b_{1}$ in
all theoretical models because it corresponds to the close to
maximal oscillator forces in closed models or dynamic
conductivities in the open models. However, only the model of an open
cascade is the most appropriate because it does not contain those
geometric configurations of an active band, inherent to the other
models, which do not ensure the optimal regime of QCL operation.

\section{Conclusions}

\begin{enumerate}

\item We developed the theory of dynamic conductivity of electrons in
open multibarrier RTS driven by the constant electric field
taking into account the interaction between electrons and high
frequency electromagnetic field.

\item  Comparing with the experimentally produced QCL having a four
barrier active band of separate cascade \cite{3} we reveal that only
the model of a complete open cascade confidently ensures the optimal
geometric design of the active band.

\item  The developed theory of dynamic conductivity of
electrons through the multibarrier RTS, after modification, can be
further used to optimize the operation of QCL, QCD
and other nanodevices by means of the choice of their geometric
design.

\end{enumerate}

\bibliographystyle{cmpj}

%

\newpage

\ukrainianpart

\title{Оптимізація роботи квантового каскадного лазера геометричним дизайном каскаду у відкритих \\ і закритих моделях}
\author{М.В. Ткач, Ю.О. Сеті, І.В. Бойко, О.М. Войцехівська}
\address{Чернівецький національний університет ім.~Ю.~Федьковича, \\
вул. Коцюбинського, 2,  58012 Чернівці, Україна}

\makeukrtitle

\begin{abstract}
\tolerance=3000%
У наближенні ефективних мас і прямокутних потенціалів розвинута
теорія електронної динамічної провідності плоскої багатошарової
резонансно-тунельної структури у постійному електричному полі в
моделі відкритої наносистеми та сил осциляторів квантових
переходів у моделі закритої системи. На прикладі експериментально
реалізованого квантового каскадного лазера з чотирибар'єрною
активною зоною окремого каскаду показано, що саме теорія
динамічної провідності у моделі відкритого каскаду найбільш
адекватно описує процес випромінювання високочастотного
електромагнітного поля при проходженні електронів крізь
резонансно-тунельну структуру у постійному електричному полі.

\keywords резонансно-тунельна наноструктура, провідність,
квантовий каскадний лазер

\end{abstract}

\end{document}